\pdfoutput=1

\documentclass{appolb}

\usepackage{amssymb}
\usepackage{hyperref}
\usepackage{graphicx}
\usepackage{amsmath}
\usepackage[numbers,sectionbib,sort&compress]{natbib}

\newcommand{\br}[1]{\langle #1\rangle}

\begin{document}

\title{Correlations with fluctuating strings%
\thanks{Talk presented by WB at {\em XXV Cracow Epiphany Conference on Advances in Heavy Ion Physics}, 8-11 January 2019}
\thanks{Supported by Polish National Science Center grant 2015/19/B/ST/00937}}

\author{Wojciech Broniowski$^{1,2}$\thanks{Wojciech.Broniowski@ifj.edu.pl}, Martin Rohrmoser$^1$\thanks{Martin.Rohrmoser@ujk.edu.pl}
\address{$^{1}$Institute of Physics, Jan Kochanowski University, PL-25406~Kielce, Poland}
\address{$^{2}$The H. Niewodnicza\'nski Institute of Nuclear Physics, \\ Polish Academy of Sciences, PL-31342~Cracow, Poland}}

\maketitle

\begin{abstract}
We present a semi-analytic approach to forward-backward multiplicity correlations in ultra-relativistic 
nuclear collisions, based on particle emission from strings with 
fluctuating end-points. We show that with the constraints from rapidity spectra, one can obtain bounds 
for the magnitude of the standard measures of the forward-backward fluctuations. The method is generic under the assumption 
of independent production from sources. For definiteness, we use the wounded quark 
model for Au+Au and d+Au collisions at the energy of $\sqrt{s_{NN}}=200$~GeV. 
\end{abstract}

\maketitle

This talk is based on our recent work~\cite{Rohrmoser:2018shp}, where more details may be found. 
Our primary goal is to understand in simple terms the mechanism of generation of the forward-backward multiplicity fluctuations in ultra-relativistic 
nuclear collisions. The presented approach generalizes in a natural way the analysis of~\cite{Broniowski:2015oif}, where only one-end fluctuations 
of strings were incorporated. 

The QCD-motivated string models are being used all over the particle physics phenomenology. In particular, numerous and 
successful Monte Carlo codes are based on the Lund model of the string formation and 
decay (see, e.g.,~\cite{Andersson:1983ia,Wang:1991hta,Lin:2004en,Sjostrand:2014zea,Bierlich:2018xfw,Ferreres-Sole:2018vgo}), or on the 
Dual Parton Model involving Pomeron and Regge exchange~\cite{Capella:1992yb,Werner:2010aa,Pierog:2013ria}. A shared feature is the formation of 
a collection of strings in the early stage of the collision. The string end-points span the color field and 
have opposite color charges (triplet with the quark-diquark or quark-antiquark, or octet with the gluon-gluon strings). The 
location of the end-points in spatial rapidity fluctuates randomly according to appropriate parton distribution functions, which together with other
incorporated effects  (such as the nuclear shadowing or baryon stopping) leads to predictions of the one-body spectra and the 
forward-backward event-by-event fluctuations. 

On the other hand, multiplicity in hadron production is successfully described within the wounded picture~\cite{Bialas:1976ed}, 
where the Glauber model~\cite{Glauber:1959aa} is adopted to describe inelastic collisions~\cite{Czyz:1969jg}. 
It has been found that the scaling based on wounded quarks~\cite{Bialas:1977en,Bialas:1977xp,Bialas:1978ze,Anisovich:1977av} rather than nucleons
works surprisingly well~\cite{Eremin:2003qn,KumarNetrakanti:2004ym,Bialas:2006kw,Bialas:2007eg,Alver:2008aq,%
Agakishiev:2011eq,Adler:2013aqf,Loizides:2014vua,Adare:2015bua,%
Lacey:2016hqy,Bozek:2016kpf,Zheng:2016nxx,Sarkisyan:2016dzo,Mitchell:2016jio,Chaturvedi:2016ctn,Loizides:2016djv,Tannenbaum:2017ixt} at
RHIC and the LHC. 

The approach of~\cite{Broniowski:2015oif,Rohrmoser:2018shp} used here merges the two above concepts in the following manner: 
\begin{enumerate}
 \item Each wounded quark pulls a string;
 \item The end-points of the strings are generated from  appropriate distributions in such a way that the experimental one-body 
 pseudorapidity spectra, serving as a constraint, are reproduced;
 \item The emission of particles from a string between its end-points is uniform in space-time rapidity;
 \item The strings emit particles independently of one-another;
 \item For clarity, we consider only one type of strings, which leads to a simple semi-analytic analysis.  
\end{enumerate}

Modeling the rapidity spectra in the adopted approach is based on the key formula holding in the nucleon-nucleon ($NN$) center-of-mass frame (CM),
\begin{eqnarray}
\frac{dN}{d\eta} = \br{N_A} f(\eta)+\br{N_B} f(-\eta), \label{eq:woundeta}
\end{eqnarray}
where at a given collision energy $f(\eta)$ should be a universal (i.e., independent of the number of participants) emission profile of a string 
pulled by a wounded quark
(we use the convention that nucleus $A$ moves to the right and $B$ to the left). 
Whereas from symmetric collisions ($A=B$) one can only obtain the symmetric part of $f(\eta)$, as then
$\br{N_A}=\br{N_B}$, from asymmetric collisions one can also extract
the antisymmetric component~\cite{Bialas:2004su,Barej:2017kcw,Adare:2018toe}. It has been found that 
$f(\eta)$ is a widely spread function in essentially the whole available range of $\eta$, with a broad peak in the forward direction. 
A simplified triangular shape of $f(\eta)$ has been used in several works~\cite{Adil:2005bb,Bozek:2010bi,Bozek:2013uha,Monnai:2015sca,Chatterjee:2017mhc}.

\begin{figure}
\begin{center}
\includegraphics[width=0.67\textwidth]{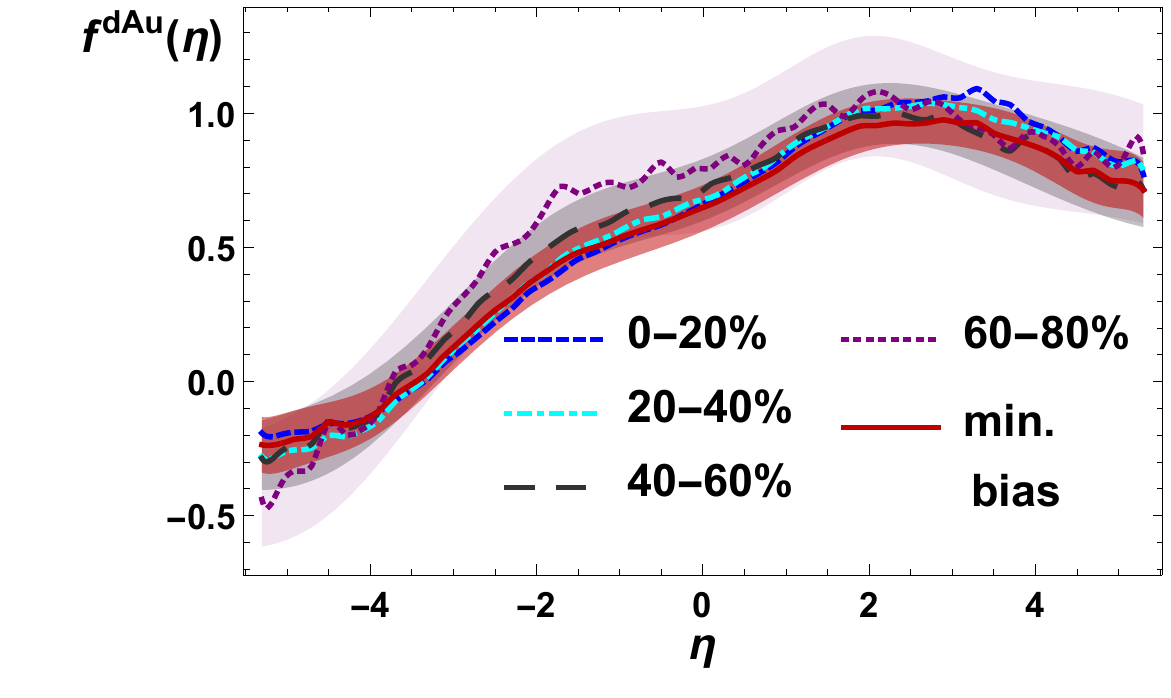}
\end{center}
\vspace{-7mm}
\caption{Emission profiles of strings pulled by the wounded quarks, obtained from fits to the pseudorapidity 
spectra in d-Au collisions at $\sqrt{s_{NN}}=200$~GeV from the PHOBOS Collaboration~\cite{Back:2003hx,Back:2004mr}.
The bands indicate the experimental uncertainties. \label{fig:dAu}}
\end{figure}

We use the original method of~\cite{Bialas:2004su} (recently applied also to the wounded quarks~\cite{Barej:2017kcw})
to extract $f(\eta)$ of Eq.~(\ref{eq:woundeta}) from the PHOBOS experimental data~\cite{Back:2003hx,Back:2004mr,Back:2002wb}.
The needed valence quark multiplicities $\br{N_A}$ and $\br{N_B}$ in the specified centrality classes were obtained from 
{\tt GLISSANDO}~\cite{Rybczynski:2013yba}. 
We note that the results for various centrality classes basically overlap within the experimental uncertainties,  
hence we may conclude that the approach yields a universal profile function $f(\eta)$ which reproduces the PHOBOS rapidity spectra 
at $\sqrt{s_{NN}}=200$~GeV. We thus confirm the results of~\cite{Barej:2017kcw}.

From the point of view of QCD, the roughly triangular shape of $f(\eta)$ seen in Fig.~\ref{fig:dAu} finds its motivation in string models, where 
one end-point of the string is associated to a valence parton, whereas the other one is randomly generated along the 
space-time rapidity and is associated to a wee parton~\cite{Brodsky:1977de}. 
Note that linking the string to a leading (wounded) parton is very much in the spirit of the Lund model~\cite{Andersson:1983ia}. 
We thus have in each event $N_A$ and $N_B$ ``wounded strings'' associated to wounded valence quarks in nuclei $A$ and $B$. 

Next, following our recent work~\cite{Rohrmoser:2018shp} we show how the end-point fluctuations 
relate to the emission profile. Let the emission of a particle with pseudorapidity $\eta$ from the string fragmentation process 
be uniformly distributed along the string between its end-points at $y_1$ and $y_2$.  Then
the emission probability distribution is 
\begin{eqnarray}
s(\eta;y_1,y_2)=\omega\left[\theta(y_1<\eta<y_2)+\theta(y_2<\eta<y_1)\right],
\end{eqnarray}
where $\omega$ is the production rate. After a short calculation we find that
\begin{eqnarray}
f(\eta) =\int_{-y_b}^{y_b} \!\!\! dy_1\,g_1(y_1) \int_{-y_b}^{y_b}\!\!\! dy_2\, g_2(y_2) s(\eta,y_1,y_2)= 
 \omega \left [ \tfrac{1}{2} - 2 H_1(\eta)  H_2(\eta)  \right ], \label{eq:f1G} 
\end{eqnarray}
with the shifted cumulative distribution function defined as
\begin{eqnarray}
H_{i}(\eta)=G_{i}(\eta)-\tfrac{1}{2}, \;\;\; G_i(\eta)=\int_{-\infty}^\eta dy \, g_{i}(y), \;\; i=1,2,
\end{eqnarray}
where $G_i(\eta)$ are the standard cumulative distribution functions (CDFs) of the string-end points.
Of course, $g_i(\eta) =dG_i(\eta)/d\eta=dH_i(\eta)/d\eta$.

It is clear from Eq.~(\ref{eq:f1G}) that the procedure of extracting $H_1(\eta)$ and $H_2(\eta)$ when $f(\eta)$ is known is not unique, as the product of two 
unknown functions is related to a known function.
Nevertheless, we will show that interesting bounds may be determined in the considered mathematical problem, since $H_1(\eta)$ and $H_2(\eta)$
are monotonic, continuous, and grow from $-\tfrac{1}{2}$ to $\tfrac{1}{2}$ as $\eta$ increases in its domain.
Thus far to the left, where $H_1(\eta)= H_2(\eta) = -\tfrac{1}{2}$, and far to the right,  where $H_1(\eta)= H_2(\eta) = \tfrac{1}{2}$,
we have $f(\eta)=0$, as it should be. In between, there must exist somewhere the zeros  $\eta^{(1)}_0$ and $\eta^{(2)}_0$ of $H_1(\eta)$ and $H_2(\eta)$, 
respectively. At these special points, as immediately follows from Eq.~(\ref{eq:f1G}), 
$f(\eta^{(i)}_0) =\tfrac{1}{2}\omega$, which is also the lowest possible value for the maximum of $f(\eta)$. 
We make a technical assumption here that 
$f(\eta)$ is unimodal, i.e., has a single maximum at $\eta_{\rm max}$ (this assumption is justified phenomenologically by the PHOBOS data). 
The situation where  $f(\eta_{\rm max})=\tfrac{1}{2}\omega$ corresponds to the special case
$\eta^{(1)}_0=\eta^{(2)}_0=\eta_{\rm max}$, where we can assume equal distributions for the two end-points:
\begin{eqnarray}
H_1(\eta)=H_2(\eta)=\sqrt{\frac{1}{4}- \frac{1}{2\omega} f(\eta)}\,{\rm sgn}(\eta-\eta_{\rm max}), \label{eq:case1}
\end{eqnarray}
We label this case ``$g_1=g_2$''. 

On the other hand, when the maximum of $f(\eta)$ is $\omega$ (the highest possible value, assumed when $H_1(\eta_{\rm max})=-H_2(\eta_{\rm max})=
-\tfrac{1}{2}$), one may choose
\begin{eqnarray}
H_1(\eta)&=& - \frac{1}{2} \theta(\eta_{\rm max}-\eta) +\left [ \frac{1}{2} - \frac{1}{\omega} f(\eta) \right ] \theta(\eta-\eta_{\rm max}), \nonumber \\
H_2(\eta)&=& - \left [ \frac{1}{2} - \frac{1}{\omega} f(\eta) \right ] \theta(\eta_{\rm max}-\eta)  + \frac{1}{2} \theta(\eta-\eta_{\rm max}).  \label{eq:case2}
\end{eqnarray}
The supports of $g_1(\eta)$ and $g_2(\eta)$ are disjoint, since $H_1$ is flat for $\eta<\eta_{\rm max}$ and 
 $H_2$ is flat for $\eta>\eta_{\rm max}$. Thus we term this case ``disjoint''.

Finally, there are intermediate cases for $\omega/2 < f(\eta_{\rm max}) < \omega$.
For instance, one may take a given form of $H_1(\eta)$ and 
then adjust $H_2(\eta)$  to satisfy  Eq.~(\ref{eq:f1G}), namely
\begin{eqnarray}
H_2(\eta)=\frac{\frac{1}{4}- \frac{1}{2\omega} f(\eta)}{H_1(\eta)}. \label{eq:h2}
\end{eqnarray}
As $-\tfrac{1}{2} \le H_1(\eta) \le \tfrac{1}{2}$,  flipping the sign at $\eta_0^{(1)}$, Eq.~(\ref{eq:h2}) yields
\begin{eqnarray}
H_2(\eta) \ge \frac{1}{2}- \frac{1}{\omega} f(\eta) {\rm ~~for~} \eta \ge \eta_0^{(1)}, \nonumber \\
H_2(\eta) \le -\frac{1}{2}+\frac{1}{\omega} f(\eta) {\rm ~~for~} \eta \le \eta_0^{(1)} \label{eq:h3}
\end{eqnarray}
(and symmetrically for $H_1$), hence formulas (\ref{eq:case2}) give the upper and lower bounds for any solution.  

\begin{figure}
\begin{center}
\includegraphics[width=0.67\textwidth]{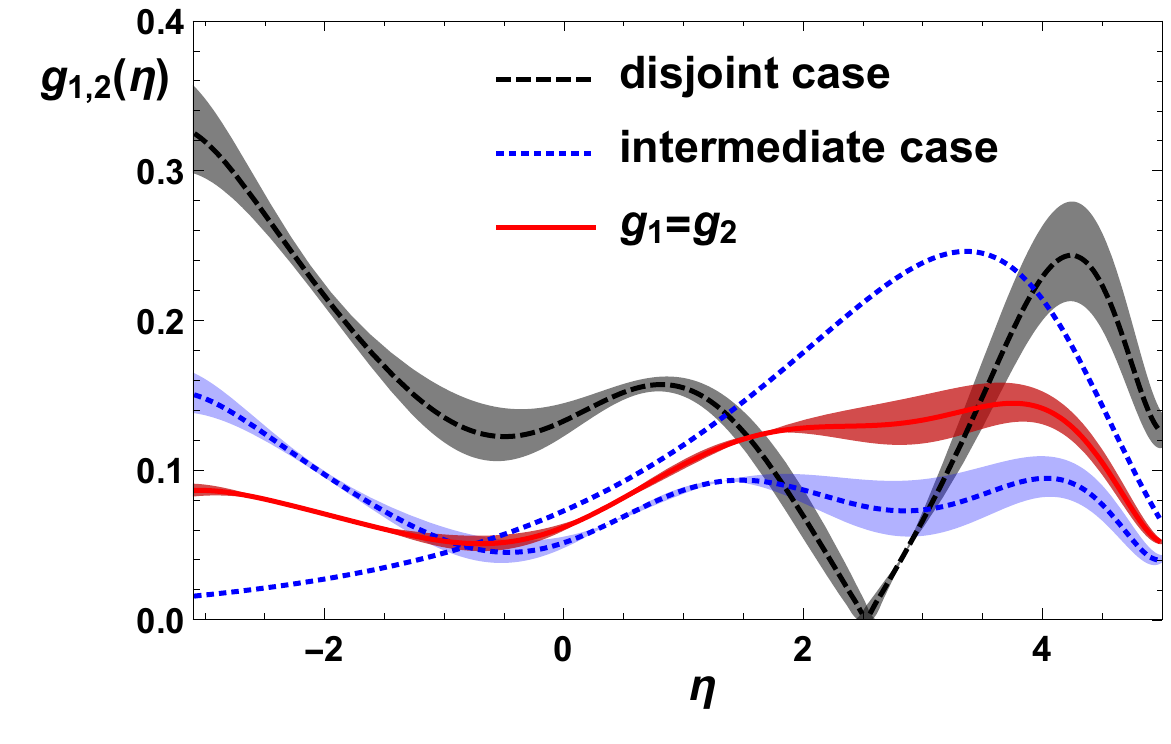}\\ \vspace{-1mm}
\includegraphics[width=0.67\textwidth]{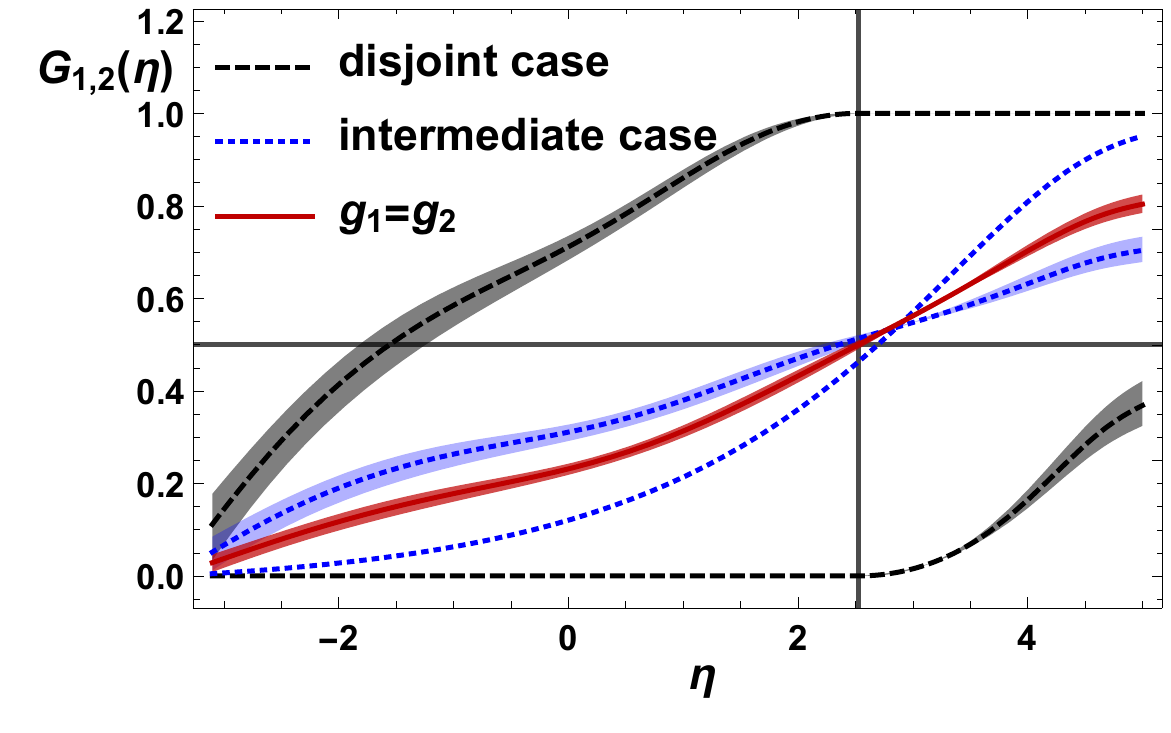}  \vspace{-7mm}
\end{center}
\caption{The end-point probability distribution functions $g_1$ and $g_2$ (top) and the 
corresponding cumulative distribution functions $G_1$ and $G_2$ 
(bottom) for various possibilities described in the text. The vertical line is placed at $\eta_{\rm max}$. \label{fig:g1g2}}
\end{figure}

Figure~\ref{fig:g1g2} presents the distributions of the string end-points and the corresponding CDFs for the three cases: $g_1=g_2$, disjoint, 
and intermediate, where one end-point is distributed according to a suitable valence quark distribution function~\cite{Rohrmoser:2018shp}.
The  bands provide uncertainties propagated from the experimental errors.
For $g_1=g_2$ (single solid line), the distribution peaks at forward rapidity (i.e., the Au side), 
as expected from the shape of the one-body profile $f(\eta)$ in Fig.~\ref{fig:dAu}. In the disjoint case 
(pairs of dashed lines), the supports for $g_1$ and $g_2$ are separated. In the intermediate case the dotted curve corresponds to the valence quark. It
is peaked in the forward direction, as expected. 
With the used parametrization of the valence quark distribution~\cite{Rohrmoser:2018shp}, the intermediate case is not 
far from the $g_1=g_2$ case.
We have verified that this holds for other  parameterizations of the valence quark parton distribution functions.
We remark that all the substantially different cases of Fig.~\ref{fig:g1g2} reproduce by construction the
``experimental'' emission profile $f(\eta)$.

\begin{figure}
\begin{center}
\includegraphics[width=0.67\textwidth]{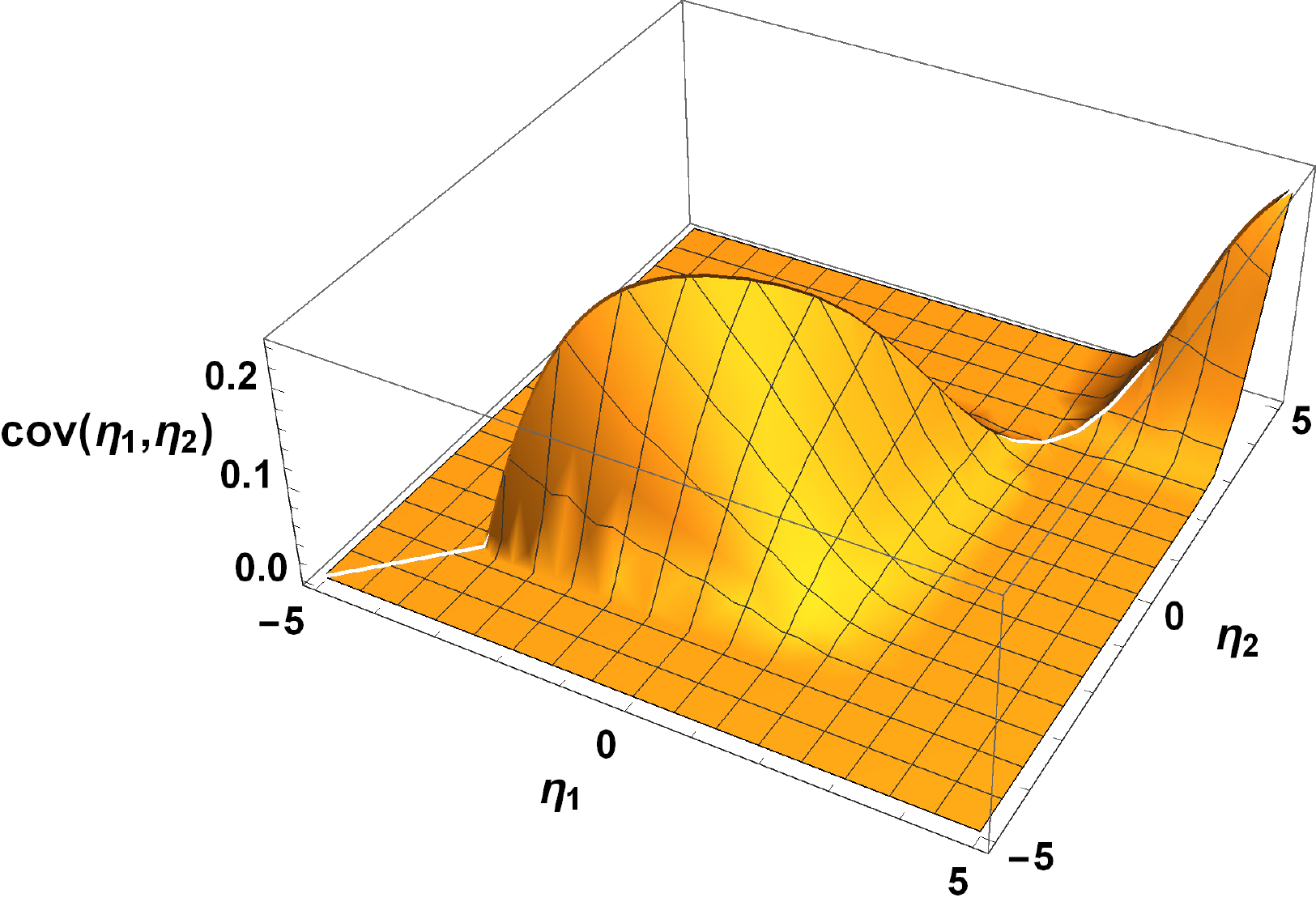}\\
\includegraphics[width=0.67\textwidth]{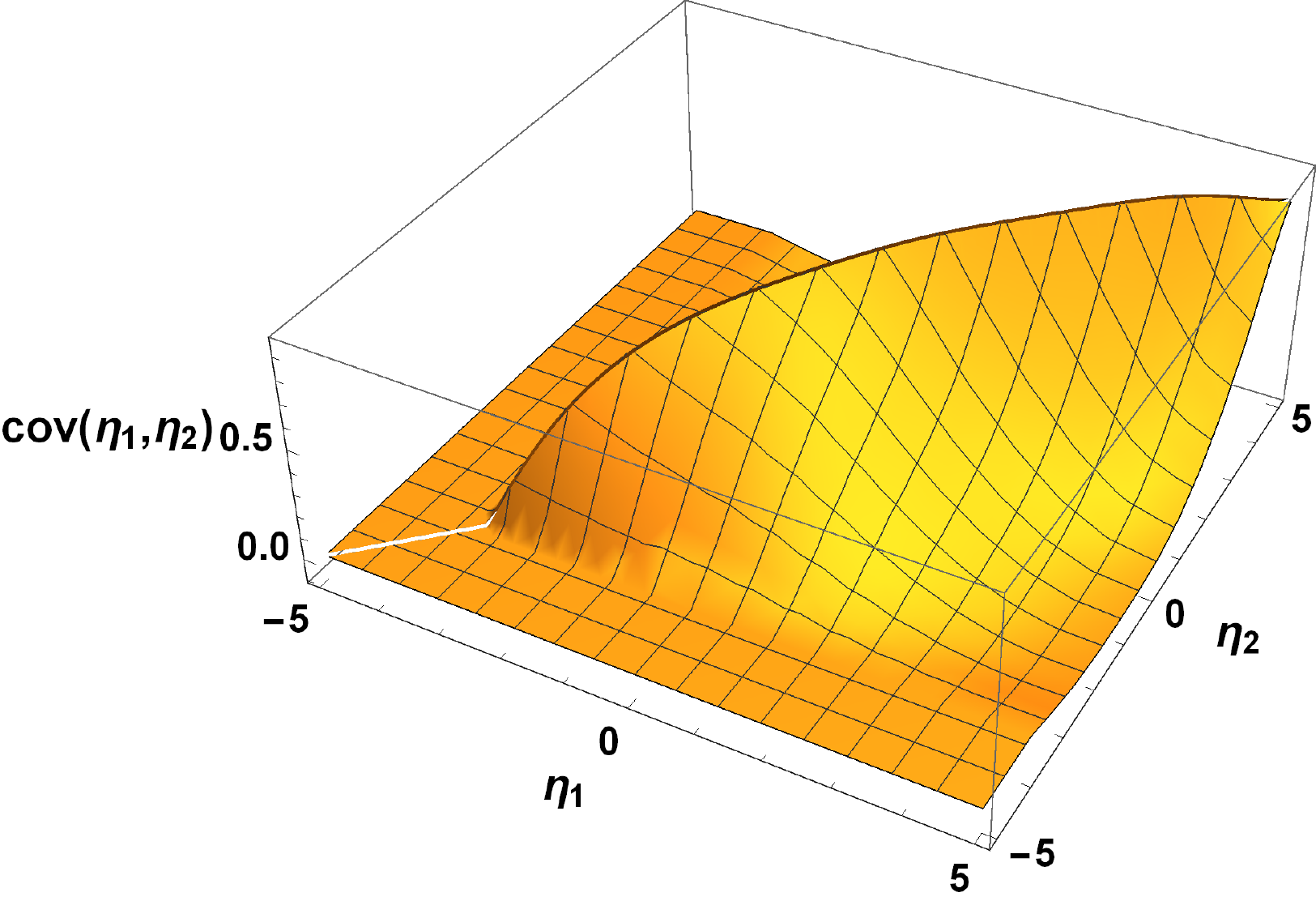}
\end{center}
\vspace{-2mm}
\caption{Covariance of the emission from a single string for the disjoint (top) and $g_1=g_2$ (bottom) cases.  \label{fig:cov}}
\end{figure}

We are ready to pass to the two-particle distributions, which is the main subject of this talk.  We have now for the two-body 
probability distribution the formula~\cite{Rohrmoser:2018shp}
\begin{eqnarray}
f_2(\eta_1,\eta_2) = \omega^2  \label{eq:fnG} G_1[{\rm min}(\eta_1,\eta_2)] \left \{1-G_2[{\rm max}(\eta_1,\eta_2)] \right \} + (1 \leftrightarrow 2).
\end{eqnarray}
The covariance of the emission from a single string is defined in a standard way as
\begin{eqnarray}
{\rm cov}(\eta_1,\eta_2)={f_2(\eta_1,\eta_2)} -  {f(\eta_1)} {f(\eta_2)}. \label{eq:cov} 
\end{eqnarray}
It is displayed in Fig.~\ref{fig:cov} for the disjoint and $g_1=g_2$ cases, which are widely different in shape as well as in magnitude, 
with the former significantly smaller than the latter.
The covariance in the intermediate case (not shown) is very close to the $g_1=g_2$ case.

In a nuclear collision, a collection of strings is formed; they ``belong'' either to the valence 
quarks from nucleus A or  B. 
With the key assumption of independent emission from different strings, the expressions for the 
one- and two-body distributions  account for simple combinatorics. For the one-body density 
in A-B collisions one has for the whole system (cf. Eq.~(\ref{eq:woundeta}))
\begin{eqnarray}
{f_{AB}(\eta)} = \br{N_A}  {f_A(\eta)} + \br{N_B}  {f_B(\eta)}, \label{eq:onebody}
\end{eqnarray}
where $f_A(\eta)=f(\eta)$ and $f_B(\eta)=f(-\eta)$, as A moves forward and B backward in the $NN$ CM frame.
Analogously, 
\begin{eqnarray}
 {\rm cov}_{AB}(\eta_1, \eta_2)  
                            &=& \br{N_A} {\rm cov}_A(\eta_1,\eta_2) + \br{N_B} {\rm cov}_B(\eta_1,\eta_2)   \label{eq:gen} \\
                            &+& {\rm var}(N_A) {f_A(\eta_1)}{f_A(\eta_2)} + {\rm var}(N_B) {f_B(\eta_1)}{f_B(\eta_2)} \nonumber \\
                            &+& {\rm cov}(N_A,N_B) \left [{f_A(\eta_1)} {f_B(\eta_2)}+ {f_B(\eta_1)} {f_A(\eta_2)} \right ]. \nonumber
\end{eqnarray}
We also introduce the customary correlation $C$ defined as 
\begin{eqnarray}
C_{AB}(\eta_1,\eta_2) =1+ \frac{{\rm cov}_{AB}(\eta_1,\eta_2)}{f_{AB}(\eta_1)f_{AB}(\eta_2)}, \label{eq:C}
\end{eqnarray}
and the $a_{nm}$  coefficients~\cite{Bzdak:2012tp,ATLAS:2015kla,ATLAS:anm}
\begin{eqnarray}
a_{nm} &=& \frac{\int_{-Y}^Y {d \eta_1} \int_{-Y}^Y {d \eta_2} C(\eta_1,\eta_2)
T_n\left(\frac{\eta_1}{Y}\right) T_m\left(\frac{\eta_2}{Y}\right)}
{\int_{-Y}^Y {d \eta_1} \int_{-Y}^Y {d \eta_2} C(\eta_1,\eta_2)}. \label{eq:anmC}
\end{eqnarray}
Here $[-Y,Y]$ denotes the covered pseudorapidity range, where for RHIC we use $Y=1$, whereas
$T_n(x)=\sqrt{n+1/2}P_n(x)$ ($P_n(x)$ denote the Legendre polynomials).

Equation~(\ref{eq:gen}) contains terms with two different kinds of fluctuations: those originating from the string end-point fluctuations, with
${\rm cov}_i(\eta_1,\eta_2)$, and terms with moments of fluctuations of the numbers of wounded quarks, $N_A$ and $N_B$.
We have found that the string end-point 
fluctuations largely dominate over the  $N_A$ and $N_B$ fluctuations,
which contribute  to $a_{11}$ at the level of 10-20\% only.

\begin{figure}
\begin{center} 
\includegraphics[width=0.67\textwidth]{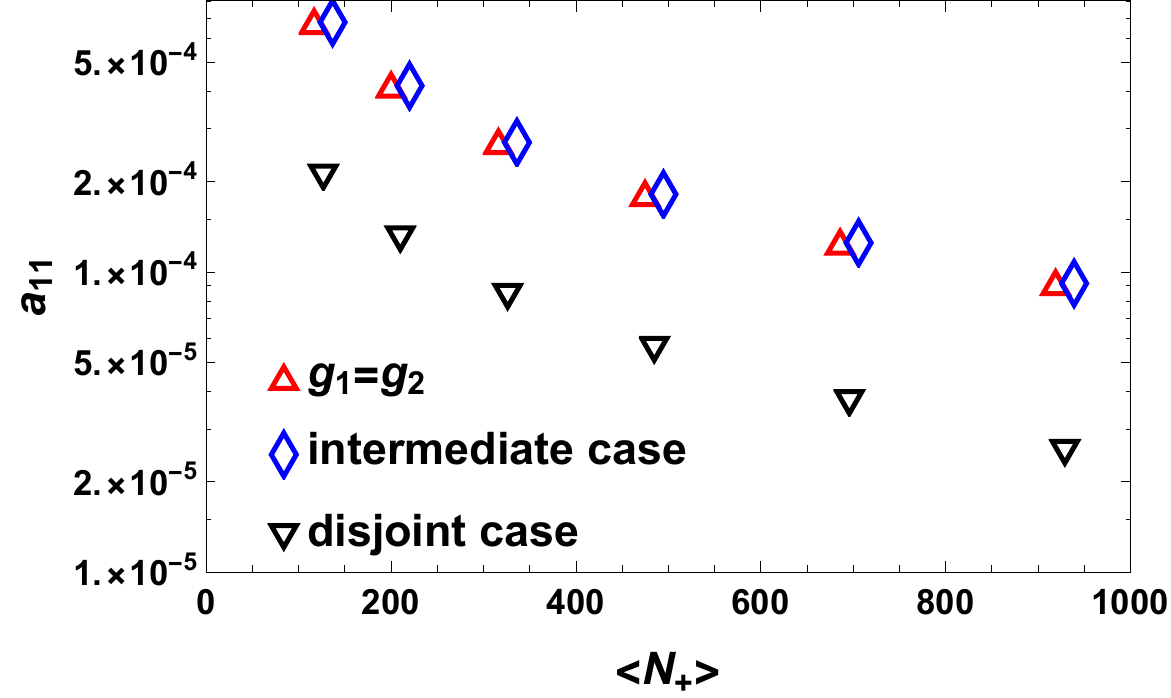} \vspace{3mm} \\
\includegraphics[width=0.67\textwidth]{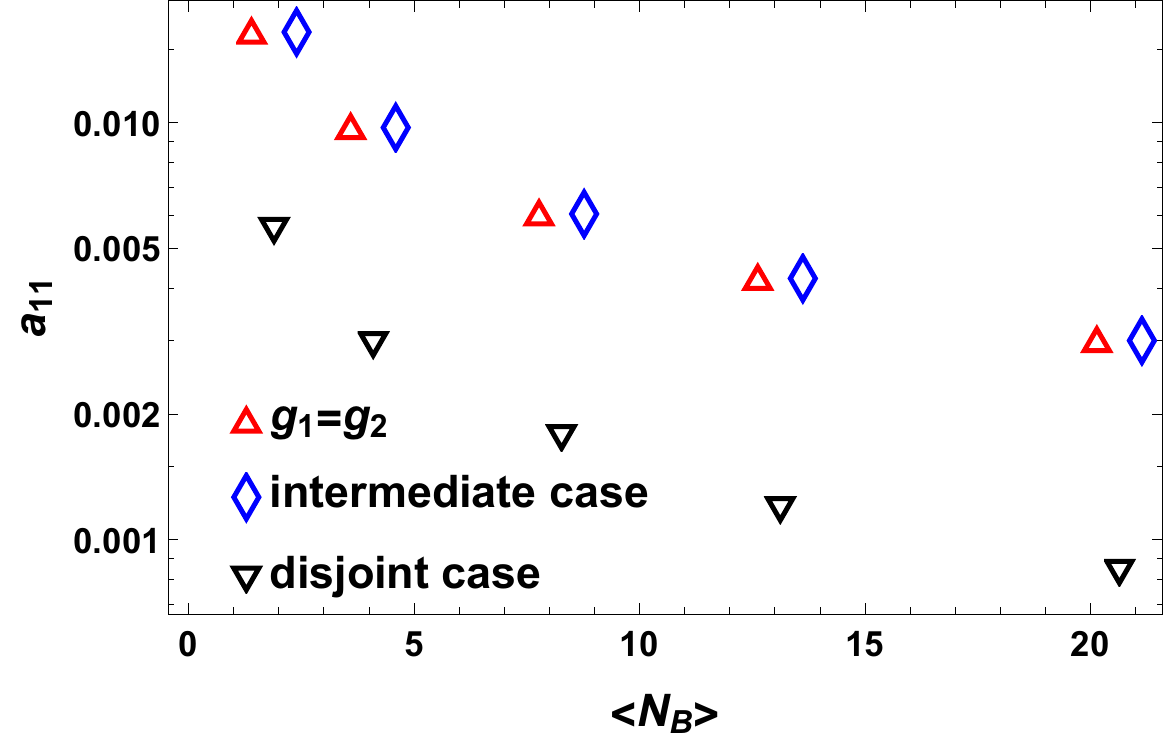}
\end{center}
\vspace{-3mm}
\caption{The predicted $a_{11}$ coefficients for Au-Au (top) and d-Au (bottom) collisions at $\sqrt{s_{\rm NN}}=200$~GeV, plotted as 
functions of $N_+=\br{N_A+N_B}$ or $\br{N_B}$, respectively.   \label{fig:a11}}
\end{figure}

Figure~\ref{fig:a11} presents the predicted $a_{11}$ for Au-Au and d-Au collisions 
at $\sqrt{s_{\rm NN}}=200$~GeV.
We note that the results for the $g_1=g_2$ and for the intermediate cases are nearly identical, while the result for 
the disjoint case is about a factor of 3 smaller. 
In view of the previous discussion, the  $g_1=g_2$ and the disjoint cases represent the upper and lower bounds. 
From Eq.~(\ref{eq:gen}) it is clear that to a good approximation  $a_{11}$ scales, as expected, with the inverse of the number of sources. 

In summary, here are our main points:

\begin{itemize}

\item Semi-analytic approach to the analysis of two-body correlations with constraint from the one-body 
pseudorapidity spectra yields nontrivial bounds for the forward-backward correlation measures;

\item Fluctuations of the string end-points yield much larger forward-backward correlations 
and dominate over fluctuations of the number of strings;

\item The $a_{nm}$ Legendre coefficients  scale as the inverse of the number of sources (strings).
 
\end{itemize}

\bibliography{hydr.bib}

\end{document}